\documentclass{jltp}
\usepackage{epsfig}

\title{Superconducting and Insulating Behavior in One-Dimensional Josephson 
Junction Arrays}

\author{David B. Haviland, Karin Andersson, Peter {\AA}gren}

\address{Royal Institute of Technology, Nanostructure Physics,SE-100 44 
Stockholm, Sweden}

\runninghead{D. B. Haviland {\it et al.} }{Superconducting and Insulating 
Behavior in One-Dimensional Josephson ...}
       
\begin{document}
\maketitle

\begin{abstract}

Experiments on one-dimensional small capacitance Josephson Junction arrays 
are described.  The arrays have a junction capacitance that is much larger 
than the stray capacitance of the electrodes, which we argue is important 
for observation of Coulomb blockade.  The Josephson energy can be tuned {\it 
in situ} and an evolution from Josephson-like to Coulomb blockade behavior 
is observed.  This evolution can be described as a superconducting to 
insulating, quantum phase transition.  In the Coulomb blockade state, 
hysteretic current-voltage characteristics are described by a dynamic model 
which is dual to the resistively shunted junction model of classical 
Josephson junctions.


\end{abstract}

\section{INTRODUCTION}

The Coulomb blockade of Cooper pair tunneling (CBCPT) is a remarkable 
phenomena, where a normally superconducting tunnel junction becomes 
insulating, behaving as a capacitor with a critical voltage for current 
flow.  The CBCPT can be observed when the Josephson coupling energy, $E_{J}$ 
is the same order of magnitude as the charging energy, $E_{C}=e^{2}/2C$, 
where $C$ is the junction capacitance.  The temperature must be low, 
$k_{B}T<E_{C},E_{J}$, and most importantly, the junction must experience a 
high impedance electromagnetic environment.  The electrodynamic environment 
controls the quantum fluctuations of the Josephson phase, and thereby 
determines whether superconducting (Josephson-like) or insulating (Coulomb 
blockade) behavior is observed.  

When the electrodynamic environment is described by the impedance 
$Z_e(\omega)$, quantum fluctuations of the phase become large if 
$Re[Z_e(\omega)]\gg R_Q$, where $R_{Q}=h/4e^{2}=6.45$k$\Omega$ is the 
quantum resistance for Cooper pairs.  Large quantum fluctuations of the 
phase mean that the number difference of Cooper pairs across the junction 
becomes a sharply defined quantum variable, as $\Delta N \Delta \phi = 1/2$ 
for a coherent state.  By realizing this high impedance environment with 
Josephson tunneling, we realize a unique state of charged matter, where many 
Bosons (Cooper pairs) are condensed into a state with well specified number.

For a single Josephson junction in an arbitrary, linear electrodynamic 
environment,  the effect of quantum fluctuations on Josephson tunneling is 
well described theoretically for the case $E_J \ll E_C$ 
\cite{ingold:rates:92}.  However, when a single Josephson junction is placed 
in the environment of several other Josephson junctions, the theory is much 
less clear.  Theoretical work on Josephson junctions arrays has primarily 
concentrated on two-dimensional (2D) arrays.  A phase diagram is often 
calculated which maps out insulating and superconducting regions, depending 
on the various parameters of the junctions, $E_J$, $E_C$, the normal state 
tunneling resistance $R_N$, or the quasiparticle tunneling resistance, 
$R_{qp}$ (for a review, see \cite{schoen:review:90}).  The theory can be 
related to experimental work on granular thin films of superconducting 
metals (for a review see \cite{katsumoto:review:95}) and 2D Josephson 
Junction Arrays fabricated with electron-beam lithography 
\cite{chen:scaling:95,zant:fieldSItrans:92}.  Finite size effects, and the 
interplay between the island stray capacitance and the junction capacitance 
are generally not considered in the theoretical treatment of Josephson 
junction arrays.  However, experiments on 2D arrays show a connection 
between the array size and the character of the superconductor-insulator 
(S-I) transition \cite{chen:arraysA:91}.  The Delft group has also studied 
long, two-dimensional arrays, which approach a 1D parallel array 
\cite{oudenaarden:quantVort:98}, as well as short 1D series arrays 
\cite{geerligs:pairtunneling:90,oudenaarden:thesis:97}

In this work we will examine experimental results on one-dimensional (1D) 
arrays of small capacitance Josephson junctions.  We will describe the 
evolution from superconducting to insulating behavior in the arrays, which 
qualitatively is very similar to that observed in granular thin films and 2D 
Josephson junction arrays.   We argue that the ratio of the junction 
capacitance to the island stray capacitance is an important parameter for 
the coupling to a dissipative environment, and therefore is crucial for 
observation of the S-I transition.

\subsection{One Dimensional Arrays}

Theoretical work on 1D arrays of small capacitance Josephson junctions has 
analyzed the charge dynamics of 1D arrays in the Coulomb blockade state 
\cite{odintsov:1DJJarraySup:96,hermon:Solitons:96,haviland:CooperPairChargeS
ol:96,p
ekola:adiabaticCP:99}. Other theoretical analysis has centered on the 
questions of determining a phase diagram, and for what values of the array 
parameters an insulating phase (Coulomb blockade) or Superconducting phase 
(Josephson effect) can be expected 
\cite{Bradley:Qfluct1D:84,cha:universalSItrans:91,sondhi:QuantPhaseTrans:96,
glazman
:newphase1D:97}.  Theoretical analysis of small capacitance 1D arrays 
usually begins with simplification of  the charging energy.  In many cases 
one takes only "on site" charging, where only the stray capacitance of each 
electrode to ground, $C_0$ is considered.  A better approximation to the 
experiments is to reduce the capacitance matrix to a bi-diagonal form, where 
$C_0$ and the junction capacitance, $C$ are taken into account.  In 
experiments, typically $C > C_0$.  In our experiments, we strive for $C \gg 
C_0$ by making the junction capacitance fairly large ( 1 to 3 fF) and the 
spacing between junctions very small ($0.2\mu$m) which leads to a smaller 
$C_0$.

When $C>C_{0}$ the electrostatic screening length extends over several 
junctions.  The potential due to one excess charge decays exponentially with 
the characteristic length $\Lambda = \sqrt{C/C_0}$.  This potential profile 
is known as a charge soliton \cite{likharev:transmissionline:89} because the 
profile moves with the tunneling charge.  Reducing the full capacitance to 
only $C$ and $C_0$ is valid only when a ground plane is located closer to 
the array than the junction spacing.  One sample (B6-22 series described 
below) had a ground plane located $1.5 \mu$m from the array, and other 
samples had no ground plane.  Thus the bi-diagonal capacitance matrix is 
only approximate.  In the absence of a ground plane, the potential 
distribution will decay less steeply than exponential, and go as $1/r$ at 
large distance.  The  charge soliton actually spreads out, and a slightly 
more effective polarization of the array occurs over the same characteristic 
distance, $\sqrt{ C/C_0}$ \cite{likharev:linearArray:95}.

\begin{figure}
\centerline{\epsfig{file=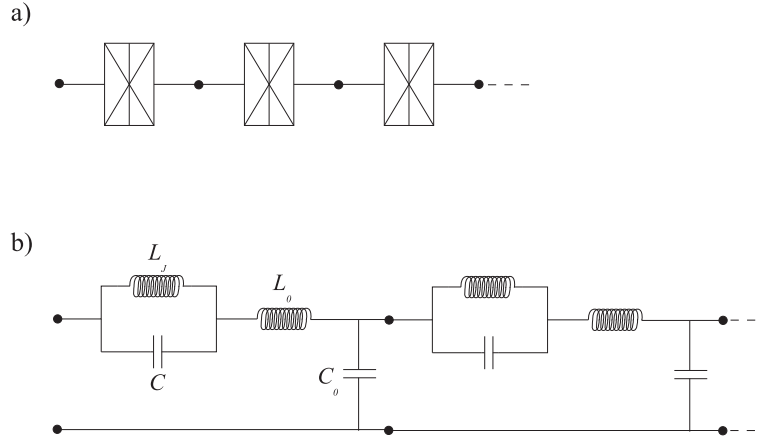,width=0.8\columnwidth}}
\caption{a) A schematic of the small capacitance Josephson junction array. 
b) An equivalent lumped element network when the phase difference across 
each junction is small.}  
\label{fig1} 
\end{figure}

\begin{figure}
\centerline{\epsfig{file=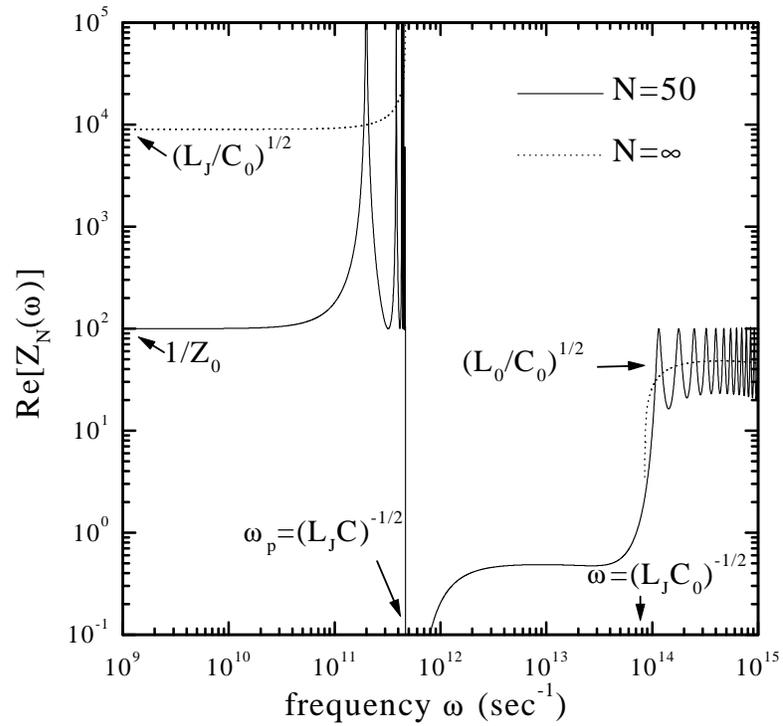,width=0.8\columnwidth}}
\caption{The impedance of the network shown in fig. \ref{fig1} calculated in 
the limit $C \gg C_0$ for the parameters of sample B6-21 (see table 
\ref{table1}):  $C=3.5$x$10^{-15}$ F, $L_J=1.1$x$10^{-9}$, 
$C_0=1.6$x$10^{-17}$, and $\sqrt{L_0/C_0}=50 \Omega$.}  
\label{fig2} 
\end{figure}

Not only the electrostatics, but also the electrodynamics of the array is 
effected by a large $\sqrt{C/C_0}$ \cite{bobbert:disipativeJJchains:92}.  
For large $\sqrt{C/C_0}$ the junctions become strongly coupled, and the 
dynamics is dominated by Josephson plasmons.  To demonstrate this point we 
consider an infinite 1D array of Josephson junctions.  For small phase 
gradient, when the current is less than the critical current, $I \ll E_J$, 
the Josephson phase-current relation can be linearized, and each junction is 
an effective inductance, $L_J=\Phi_0/2\pi I_c$, where $\Phi_0=h/2e$ is the 
flux quantum.  Replacing each Josephson junction by a parallel combination 
of $L_J$ and $C$, we may calculate the impedance of the 1D  network shown in 
fig \ref{fig1}.  The results of such a calculation are shown in fig. 
\ref{fig2} for an infinite array, and a finite array with $N=50$ junctions, 
for typical values of $I_c$, $C$ and $C_0$ in our experiments.  

For a finite array, we see a set of discrete resonances which are the 
Josephson plasmon modes.  The width of these resonances is determined by the 
terminating impedance, $Z_0$.  For the infinite array, these resonances form 
a dense set of modes, where the impedance is much larger than $Z_0$. At 
frequencies much less than the Josephson plasma frequency, $\omega _p=\sqrt 
{8E_JE_C}$ and for $C\gg C_0$ the real impedance can be written as,

$$Z_A(\omega )=\sqrt {{{L_J} \over {C_0}}}=R_Q\sqrt {{{4E_C} \over 
{E_J}}}\sqrt {{C \over {C_0}}}\ \ \ ,\ \ \omega <\omega _p={{\sqrt 
{4E_JE_C}} \over \hbar }$$

To observe the CBCPT, it is necessary that $E_C\sim E_J$ and both 
$E_C,E_J\gg k_BT$.  If $E_C\gg E_J$ the maximum Cooper pair current ("Zener 
current" ) is suppressed exponentially and becomes immeasurable, and if 
$E_C\ll E_J$ the threshold voltage is exponentially suppressed and becomes 
immeasurable \cite{likharev:blochoscillations:85}.  Thus, $Z_A\gg R_Q$ 
requires that $C\gg C_0$.  This impedance does not come from electromagnetic 
fields, as in a usual transmission line, but rather due to long wave-length 
Josephson plasmons involving the phases of many junctions.

The large ratio $\sqrt{C/C_0} \sim 10$ in our arrays not only results in a 
large impedance, and therefore a Coulomb blockade, but also influences the 
effect of random offset charges.  Simulations show that the static 
background potential due to random offset charge is smoothed by the 
averaging effects of the large ratio $C/C_0$ \cite{johansson:unpub}.  One 
would expect that pinning of the charge in some local energy minimum in the 
array could be practically eliminated.  However, simulations show that {\it 
incoherent} charge transport is substantially effected by static random 
background charge, even in the limit $C>C_0$ 
\cite{johansson:unpub,Middleton:ArraysOffset:93,matsuoka:shot1Darray:98,mels
en:inho
moIDarray:97}.  To our knowledge, the effect of random offset charges on 
{\it coherent} Cooper pair tunneling has not been investigated 
theoretically.

\section{EXPERIMENTAL RESULTS}

\begin{figure}
\centering
\epsfig{file=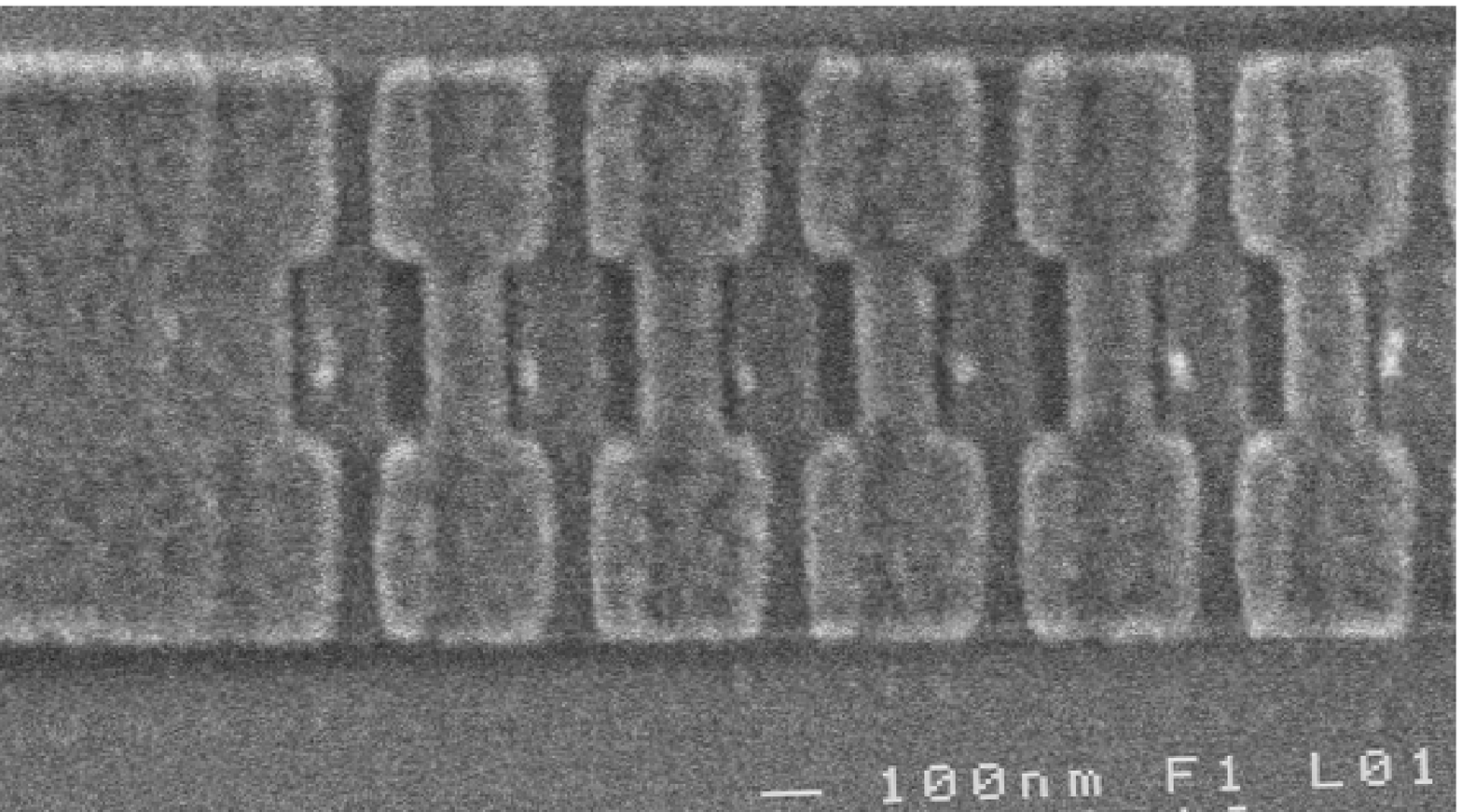,width=0.8\columnwidth}
\epsfig{file=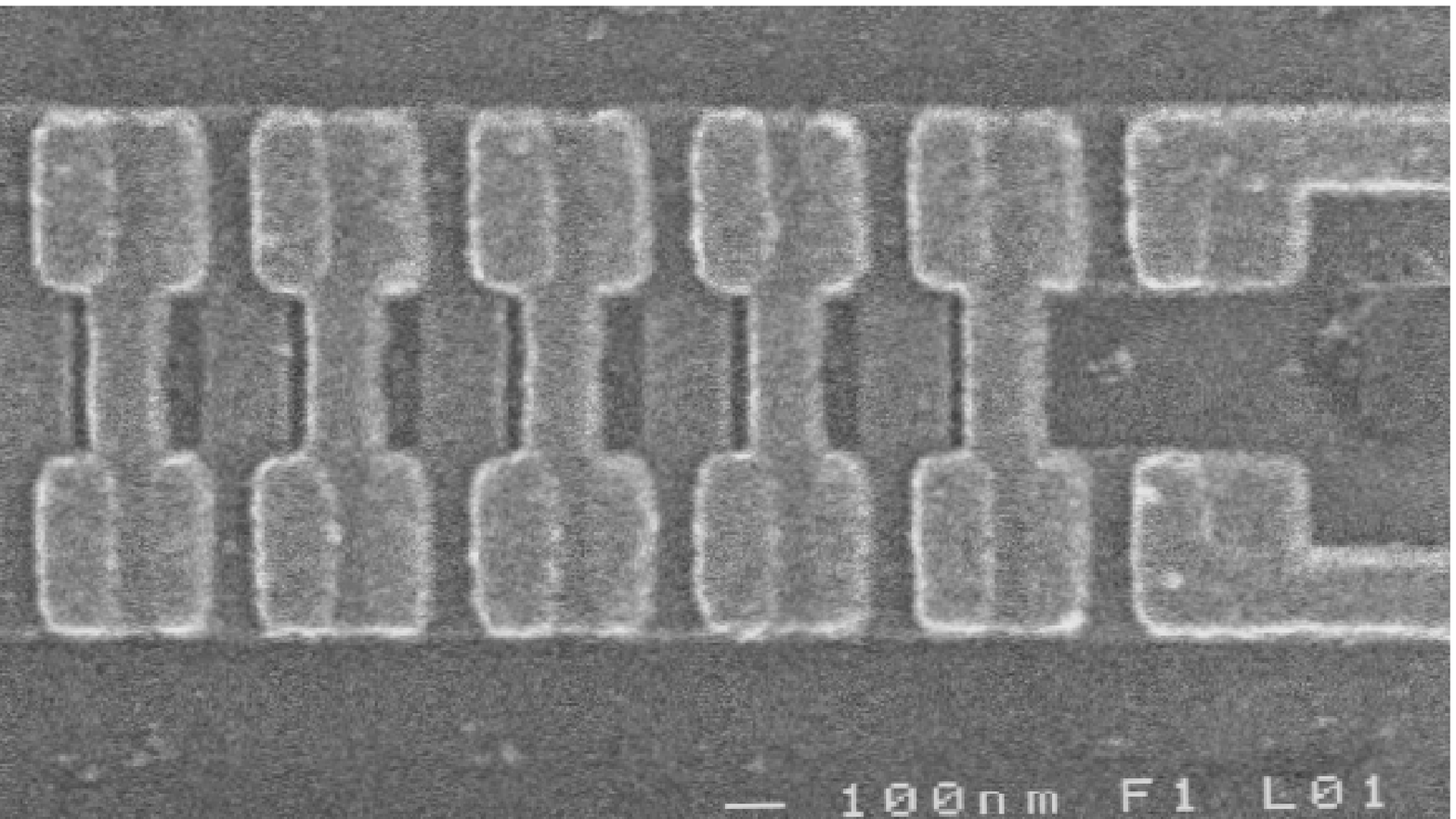,width=0.8\columnwidth}
\caption{Scanning Electron Microscope images of the arrays of SQUIDs showing 
how the array were terminated for the 2 point (upper) and 4 point (lower) 
measurement scheme.}
\label{fig3}
\end{figure}

The arrays are actually serially connected SQUIDs (Superconducting Quantum 
Interference Devices).  The SQUID geometry allows one to externally impose a 
phase shift between neighboring superconducting electrodes by application of 
a magnetic flux.  This external phase shift can be considered as effectively 
tuning the Josephson coupling energy with an external magnetic field, 
$E_{J}(B)$.  
Figure \ref{fig3} shows scanning electron micrographs of two different 
arrays.  The electrodes were Al, deposited on an oxidized Si substrate, with 
an Al$_2$O$_3$ tunnel barrier.  At the right hand side of fig.\ref{fig3} we 
see how the arrays were terminated.  Two different arrangements were used.  
Figure \ref{fig3}a shows a 2-point termination, where a wide strip connects 
to the first SQUID.  This 2-point arrangement had the disadvantage that 
expelled magnetic flux from the wide strip focused more flux into the edge 
loops.  Consequently the two edge junctions had a smaller period of 
modulation with magnetic field.  Figure \ref{fig3}b shows the 4-point 
termination, where no edge effect with the magnetic field was observed.  For 
the 4-point termination, the junctions in the leads are not SQUIDs, and thus 
do not have a tunable $E_{J}$. 

Table \ref{table1} shows the main parameters of the arrays studied thus far.  
Samples with the same chip code (B6-\#\#) were made simultaneously on the 
same chip.  The samples are arranged with decreasing $E_{J0}/E_C$.  The note 
indicates whether the array was terminated with a 2 point or 4 point lead 
configuration.  The value of $E_{J0}=(\Phi _0 / 2\pi) I_c = R_Q \Delta _0 / 
2 R_N$ was determined from the measured normal state resistance $R_N$ and 
the superconducting energy gap, $\Delta _0$.  The charging energy 
$E_C=e^2/2C$ was calculated from the observed area of the junction and the 
specific capacitance $c_s=45$fF$/\mu$m$^2$.  Depending on the array 
parameters, different behavior of the arrays could be observed.  We divide 
these into three different regions discussed below.

\begin{table}
\caption{Main parameters of studies arrays.   The notes have the following 
meaning:  2 pt. or 4 pt. describes the array termination.  SI-trans means 
that  superconductor-insulator transition could be observed.  Short arrays 
showed {\it no} Coulomb blockade of Cooper pair tunneling (CBCPT) and longer 
arrays with smaller $E_{J0}/E_C$ showed {\it only} CBCPT (no superconducting 
state)}
\label{table1}

\begin{center}
\begin{tabular}{l||l|l|l|l|l|l|l}
 Sample & $N$ & $R_N$ & $E_{J0}$ & $E_C$ & $E_{J0}/E_C$ & $\sqrt{L_J/C_0}$ & 
Note \\
 code & Jun. & (k$\Omega$) & ($\mu$eV) & ($\mu$eV) &   & (k$\Omega$) &   \\
\hline
 B6-21a & 255 & 1.1 & 586 & 23 & 25 & 8.5 & 2 pt., SI-trans \\
 B6-21b & 127 & 1.1 & 586 & 23 & 25 & 8.5 & 2 pt., SI-trans \\
 B6-21c & 63  & 1.1 & 586 & 23 & 25 & 8.5 & 2 pt., SI-trans \\
 B6-21d & 31  & 1.1 & 586 & 23 & 25 & 8.5 & 2 pt., {\it no} CBCPT \\
 B6-21e & 15  & 1.1 & 586 & 23 & 25 & 8.5 & 2 pt., {\it no} CBCPT \\
 B6-21f & 7   & 1.1 & 586 & 23 & 25 & 8.5 & 2 pt., {\it no} CBCPT \\
\hline
 B6-41a & 255 & 1.6 & 406 & 29 & 14 & 10.2 & 4 pt., SI-trans \\
\hline
 B6-32a & 255 & 2.7 & 227 & 29 & 7.8 & 13.3 & 4 pt. SI-trans at$R_Q$ \\
\hline
 B6-22a & 255 & 4.6 & 142 & 23 & 6.1 & 17.4 & 2 pt., SI-trans \\
 B6-22b & 127 & 4.6 & 142 & 23 & 6.1 & 17.4 & 2 pt., SI-trans \\
 B6-22c & 63  & 4.6 & 142 & 23 & 6.1 & 17.4 & 2 pt., SI-trans \\ 
\hline
 B6-41b & 255 & 4.3 & 149 & 50 & 3.0 & 16.8 & 4 pt., {\it only} CBCPT \\
 \hline
 B6-32b & 255 & 8.5 & 78  & 52 & 1.5  & 24 &   4 pt., {\it only} CBCPT \\
 \hline
 K9-32 & 255 & 10.4 & 62 & 59 & 1.1 & 26 & 4 pt., {\it only} CBCPT \\
 \hline
 K9-43a & 65 & 15.4 & 42 & 59 & 0.71 & 32 & 4 pt., {\it only} CBCPT \\
 K9-43b & 255 & 15.7 & 41 & 59 & 0.70 & 32 & 4 pt., {\it only} CBCPT \\

\end{tabular}
\end{center}
\end{table}

\subsection{Superconducting State}

\begin{figure}
\centering
\epsfig{file=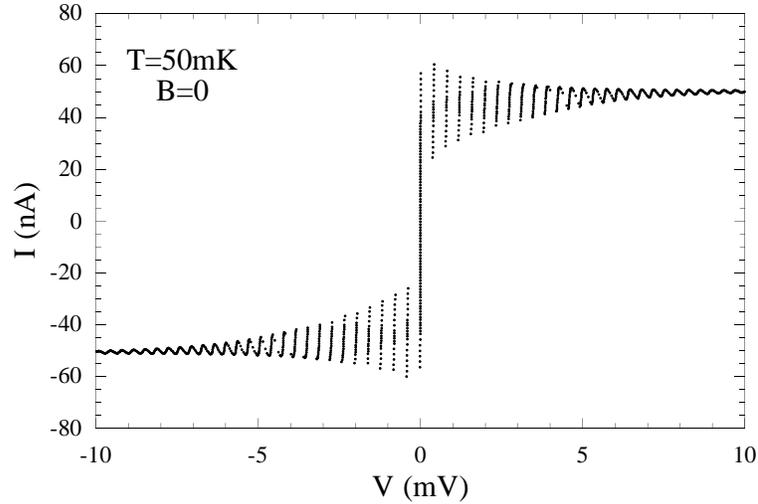,width=0.8\columnwidth}
\caption{The current voltage characteristic for sample B6-21a measured at 
$T=50$mK and $B=0$. }
\label{fig4}
\end{figure}

For the samples in table \ref{table1} with a smaller ratio $L_J/C_0$, the  
current-voltage (I-V) characteristic exhibited Josephson-like behavior at 
zero magnetic field.  Figure \ref{fig4} shows the I-V curve of sample 
B6-21a.  Here we see a nearly vertical "super current" branch of the I-V 
curve near zero voltage, and a successive set of vertical quasi-particle 
branches separated in voltage by $2\Delta _0/e$.  The dc load in the bias 
circuit was such that we could not trace out the negative differential 
resistance between these vertical quasi-particle branches. At higher 
voltages the successive quasi-particle branches merge into one flat branch 
with a current independent of voltage.  At even higher voltages, the I-V 
curve eventually switches to the normal tunneling branch.

The I-V curve shown in fig. \ref{fig4} is not that expected for arrays of 
classical, non-interacting Josephson junctions.  As the bias is increased in 
such arrays we would expect an increasing series of switching currents, 
starting from the lowest critical current in the array.  For these small 
capacitance, strongly coupled Josephson junctions, we rather observe a {\em 
decrease} of the switching current as we move out in voltage, as well as an 
increase of the retrapping current.  We have no quantitative explanation for 
this observed behavior.  We speculate that it has to do with electromagnetic 
interaction of the junctions, which are strongly coupled due to the very 
small $C_0$.  

As the magnetic field is increased, the critical current of all junctions is 
decreased.  The slope of the "critical current" branch also increases, and 
as $E_J$ is further suppressed, a distinct threshold voltage emerges as the 
Coulomb blockade of Cooper pair tunneling (CBCPT) becomes observable (see 
fig. \ref{fig5})  

\subsection{Superconductor - Insulator Transition}

\begin{figure}
\centering
\epsfig{file=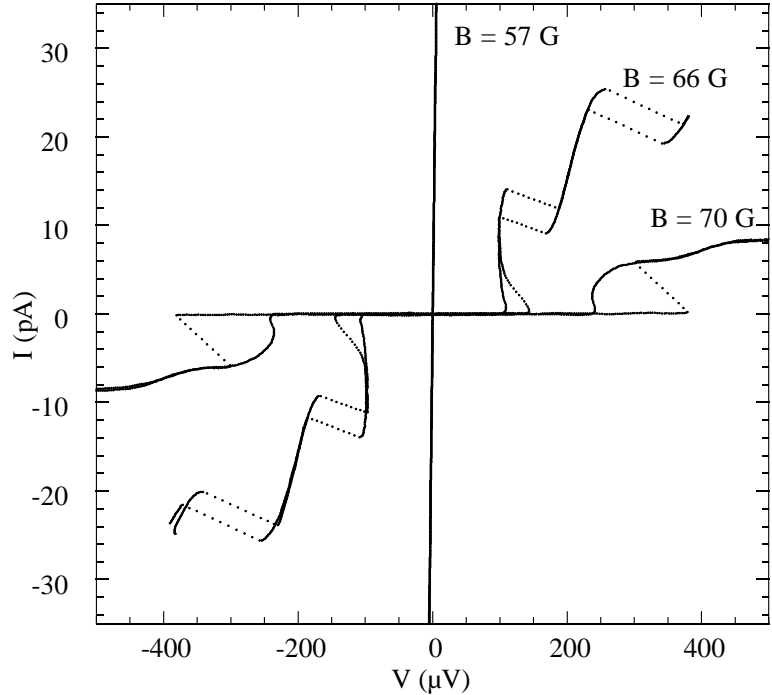,width=0.8\columnwidth}
\caption{The I-V characteristics at $T=50$mK for three different magnetic 
fields as the array B6-22a goes through the Superconductor-Insulator 
transition.  The curve labeled $57$G is actually on the insulating side of 
the transition, as $R_0$ increases when $T \rightarrow 0$ at this magnetic 
field.}
\label{fig5}
\end{figure}

Figure \ref{fig5} shows three I-V curves at different magnetic fields.  The 
nearly vertical line (I-V at $B=57$G) has a finite slope corresponding to 
about $300$ k$\Omega$.  The critical current and the hysteretic switching 
between the "supercurrent branch" and the quasiparticle branch is off scale 
for this curve.  As the magnetic field is increased the I-V curve develops a 
very high resistance state for voltages below a threshold voltage, which is 
the CBCPT.  The curve labeled $B=70$G in fig \ref{fig5} shows a hysteretic 
I-V curve in this Coulomb blockade state, meaning that "back bending" or 
region of negative differential resistance exists at low current.  We will 
discuss this hysteresis in the next section.  The intermediate curve 
($B=66$G)in fig. \ref{fig5} shows remnants of both types of hysteresis.   
  
Figure \ref{fig5} shows that the crossover to the CBCPT is characterized by 
a set of very nonlinear I-V curves.  We would like to know the {\it linear} 
response of the array as in undergoes this transition from Josephson-like to 
Coulomb blockade behavior.  To this end we have devised a measurement of the 
zero-bias resistance, 
$R_{0}\equiv \lim_{V \to 0}dV_{rms}/dI_{rms}$
under the condition that the power dissipation by the array is constant.  
$R_{0}$ is measured by phase sensitive detection of a small signal 
excitation with two lock-in amplifiers.  The excitation is adjusted so that 
the product of the two signals $(dI)_{rms}(dV)_{rms}=10^{-16}$ Watt. 

\begin{figure}
\centering
\epsfig{file=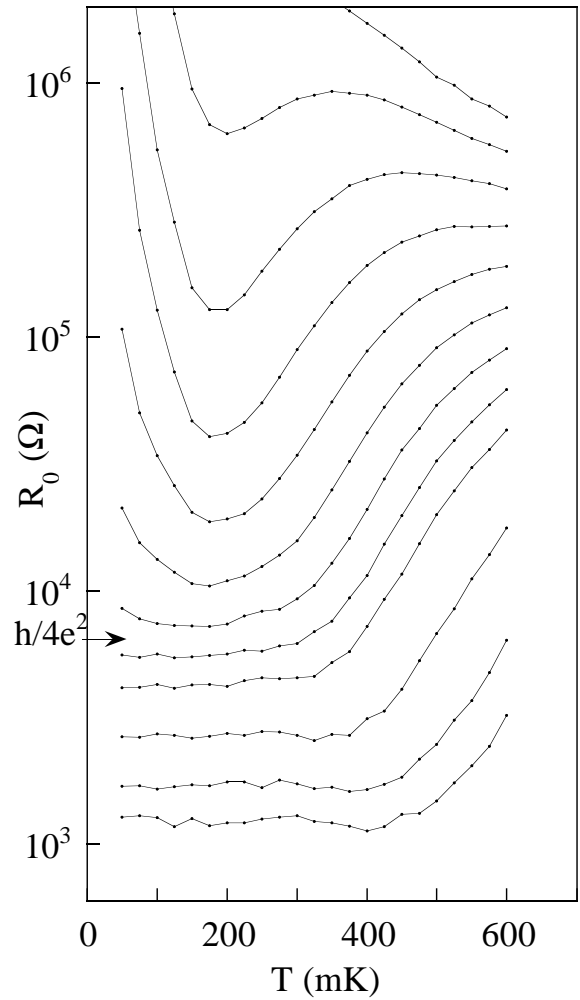,width=0.6\columnwidth}
\caption{The S-I transition is seen by measurement of the zero bias 
resistance vs. temperature.  The data are taken on sample B6-32a at several 
values of the magnetic field in the range 0 to 64 G.}
\label{fig6}
\end{figure}

$R_0$ could be measured versus temperature and magnetic field. Figure 
\ref{fig6} displays the results of such measurements.  At low magnetic 
fields (bottom curves of fig. \ref{fig6}), $R_0$ decreases as the array is 
cooled below the transition temperature of the Al ($T_C=1.2$K) and 
eventually becomes flat, showing a temperature independent resistive state 
as $T \rightarrow 0$.  The finite resistance corresponds to the slope of the 
"critical current" branch of the I-V curve (see curve $B=57$ in fig. 
\ref{fig5}).  Experiments indicate \cite{chow:SItrans1D:98} that in longer 
arrays the resistance of this "flat tail" decreases, indicating the non-zero 
resistance in this "superconducting" state is due to the finite size of the 
array.   When the magnetic field is increased, the resistance rises and the 
"flat tail" remains.  Further increasing the magnetic field causes a 
transition to a different kind of behavior, where $R_0$ increases as $T 
\rightarrow 0$, indicating the development of a Coulomb blockade. 

We have previously shown \cite{chow:SItrans1D:98} how length scaling of 
$R_{0}$ could be used to determine the critical point between 
superconducting and insulating behavior. Length scaling analysis allowed us 
to determine a $T=0$ critical point where $R_0$ is independent of length.  
The critical point  separates those curves with a "flat tail" in $R_0(T)$ 
from those with increasing
$R_0$ as $T \rightarrow 0$.  In the experiment shown in fig. \ref{fig6} we 
find that this transition occurs at $R_0(T=0)\simeq h/4e^2=6.45$k$\Omega$ 
(see the arrow in fig. \ref{fig6}). The transition does not always occur at 
$R_{Q}$, and 
the measurements displayed in fig. \ref{fig6} were special in 
that the array was terminated with a 4 point configuration.  The 4 
point termination should reduce any series measurement of dissipation 
arising at the edge of the array.  However, one junction is probably 
not enough to effectively isolate the array from the environment.  
Further experiments are needed to determine if this transition at 
$R_{0}=R_{Q}$ is universal.

\subsection{Insulating State}

\begin{figure}
\centering
\epsfig{file=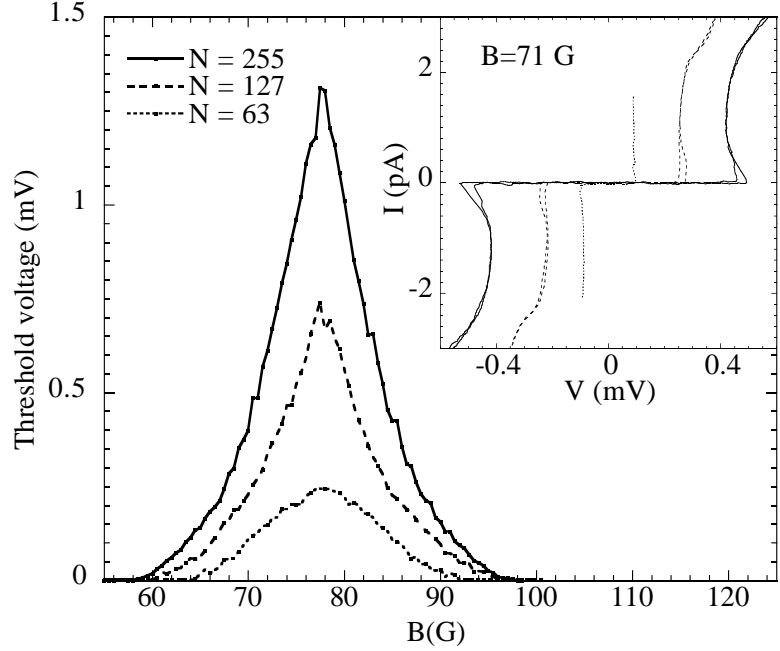,width=0.8\columnwidth}
\caption{The magnetic field dependence of the threshold voltage for sample 
B6-22. The inset shows the I-V curves. }
\label{fig7}
\end{figure}

When the CBCPT is well developed in the arrays, $R_0$ is essentially 
infinite, and we have a zero current state below the threshold voltage.  
Figure 
\ref{fig7} shows how the threshold voltage changes with magnetic field on 
the insulating side of the S-I transition.  We typically find that $V_t(B)$ 
exhibits a sharp cusp as $E_J \rightarrow 0$ when $\Phi_{ext} \rightarrow 
\Phi_0/2$ ($B=78$ G for this sample).  For samples B6-22 and B6-21 where we 
could measure length effects, we found that $V_t$ was roughly proportional 
to length.  The 
threshold voltage depends on the magnetic field in a periodic way, 
with period corresponding the flux quantum in each loop.  The 
magnitude of the current is also periodic in the magnetic field.  These 
observations prove that the current and the threshold voltage are due to 
Cooper pair tunneling.  Although the Cooper pair tunneling is a coherent 
process, we measure dissipation (finite current and non-zero voltage) 
because the pair tunneling interacts with dissipative degrees of freedom.

As in the case of a single small capacitance Josephson junction, 
dissipation arises from excitation of the electromagnetic environment 
\cite{ingold:rates:92} or quasi-particle tunneling \cite{schoen:review:90}.  
Quasi- particle tunneling becomes important when the voltage drop across any 
junction exceeds the gap voltage, $2\Delta / e = 400 \mu$V.  It is important 
to note that 
although the voltage over the entire array may be greater than $2\Delta / 
e$, this voltage is distributed over many junctions.  The charge soliton 
length for Cooper pair charge solitons is determined by an effective 
capacitance of the junction \cite{haviland:CooperPairChargeSol:96}, 
$\lambda_{s}=\sqrt{C_{eff}/C_0}=\sqrt{(2e/2\pi V_{C})/C_0}$.  The critical 
voltage $V_{C}(E_{J}/E_{C})$ goes to zero exponentially for $E_{J} \gg 
E_{C}$ \cite{likharev:blochoscillations:85}.  Hence,
$\lambda_{s} \rightarrow \infty$ at the S-I transition and near this 
transition the voltage is dropped uniformly over the 
entire array.

\begin{figure}
\centering
\epsfig{file=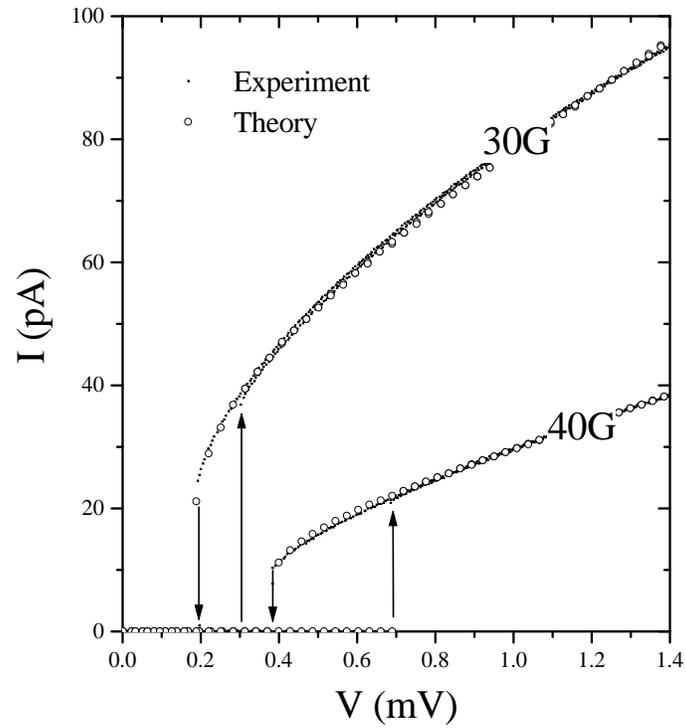,width=0.8\columnwidth}
\caption{Hysteretic I-V curves in the insulating state are observed.  The 
open circles are fits to the experimental data (dots) using the non-linear 
dynamic model for the quasicharge which is dual to the RSJ model of typical 
Josephson junctions.}
\label{fig8}
\end{figure}

In the insulating state, a very interesting model for charge transport has 
been developed, which has a direct duality to flux transport in long 
Josephson junctions \cite{benjacob:chargesoliton:89}.  A static version 
($I=0$) of this model has 
been used to explain the magnetic field dependence of the threshold 
voltage \cite{haviland:CooperPairChargeSol:96}.  We have been able to extend 
the model to the time domain, calculating the dc I-V curve and 
fitting it to the data.  This dynamic model is able to explain the 
hysteresis observed in the dc IV curve (see fig. \ref{fig8} and the curve 
labeled $B=70$G in fig. \ref{fig5}).  

In the insulating state, the Josephson junctions in the 1D chain are 
described in terms of the dimensionless quasi-charge $\chi$.  If a 
phenomenological resistance $R$ and inductance $L$ are introduced together 
with the capacitance $C_0$, a sine-Gordon like model can be derived to 
describe the dynamics of $\chi(x,t)$\cite{haviland:CooperPairChargeSol:96}.  
Soliton solutions exist for this model, and the inductive term is what gives 
the charge soliton mass, or inertia.  We may simplify the full 
sine-Gordon-like model  by dropping the second spatial derivative in order 
to capture the basics of the dynamic behavior.  The array is then viewed as 
one lumped element, which is valid if the array is shorter than the soliton 
length, as is the case near the S-I transition.  Alternatively, the second 
time and spatial derivatives may combined by assuming a traveling wave 
solution to the sine-Gordon-like model.  Either approach results in the 
following equation \cite{agren:SRJ:99}:

$${{V+RI_Z} \over {V_C}}=\beta \ddot \chi +\dot \chi + \ \ saw\ \ (\chi )$$
where $V$ is the applied voltage, and the dc current is given by the time 
average $<\dot \chi >$.  Here saw$(\chi )$  is a $2 \pi$ periodic function 
describing the charging and discharging of 
the junctions capacitance by single Cooper pair tunneling events.  The 
dot refers to differentiation with respect to dimensionless time, 
$t/RC$.  For the lumped element case, the dimensionless parameter,
$$\beta ={L \over {R^2\left( {{{2e} \over {2\pi V_C}}} \right)}}$$
determines the amount of hysteresis observed in the dc I-V curve.  We have 
modeled the dissipation arising from Zener transitions in the Bloch energy 
bands of the  single junction Coulomb blockade 
\cite{likharev:blochoscillations:85}.  To make the model simple we assumed 
that the resistance is zero below the Zener threshold current, $I_{Z}$, and 
characterized by a linear resistance $R$ above $I_Z$. 

The parameters $V_{C}$, $R$, and $I_{Z}$ are fixed for a particular magnetic 
field ($E_{J}/E_{C}$), and are determined from the observed I-V curve.  The 
parameter $\beta$ or $L$ is then adjusted to fit the observed amount of 
hysteresis.  Good fits to the data can be made, as seen in fig. \ref{fig8}.  
However, the inductance $L$ required to explain the observed hysteresis is 
extremely large, $\sim 10^{-2}$ H.  It is certainly not due to 
electromagnetic inductance ($\sim 10^{-14}$ H) and too small to be due to 
Josephson inductance ($N \Phi_{0}/2 \pi I_{C} \sim 10^{-4}$ H).  We do 
observe that the value of $L$ obtained from these fits does diverge as 
$I_{C}\rightarrow 0$ as expected for the Josephson inductance 
\cite{agren:SRJ:99}. 

Although the origin of this large inductance is not understood, we emphasize 
that an inductive term is absolutely necessary if we are to explain the 
hysteresis with any dynamic model.  We further stress that this 
hysteresis is clearly associated with the presence of Josephson 
Coupling in the arrays.  Such hysteresis has not been observed in any normal 
tunnel junction arrays to the best of our knowledge.  In our arrays, 
the normal state behavior shows no clear zero current state.  The normal 
sate is achieved at large magnetic fields ($B \sim 1000$G) where the 
superconducting energy gap is suppressed.  Only a weak remnant of the 
Coulomb blockade exists in the normal state because the normal tunnel 
junction resistance is the order of the quantum resistance.

\section{CONCLUSION}

We have given an overview of the measured transport characteristics of 1D 
small capacitance Josephson junction arrays.  The interplay between 
Josephson and Charging energies is clearly seen as a transition between 
Josephson-like to Coulomb blockade I-V curves.  Coulomb blockade is observed 
even when $E_J\geq E_C$, depending on the length of the array.  We point to 
the importance of the stray capacitance $C_0$ and the long wavelength 
Josephson plasmons in determining the coupling to dissipation.  A quantum 
phase transition can be observed by measuring the temperature dependence of 
the zero bias resistance as the effective Josephson coupling is modulated in 
the arrays.  In the insulating state, the hysteretic current voltage 
characteristics can be explained by a dynamic model which is dual to the 
usual RSJ model for Josephson junctions.

\section*{ACKNOWLEDGMENTS}
We wish to acknowledge support from the Swedish NFR, and the EU grant 22953 
CHARGE and the EU grant SMT4-CT96-2049 SETamp.  Samples were made at the 
Swedish Nanometer Laboratory.


\end{document}